\documentstyle[11pt]{article}
        \def\be{\begin{equation}}
        \def\ee{\end{equation}}

\begin{document}

\title {\normalsize \bf  Relation of The New Calogero Models and xxz Spin Chains
}

\author {V. Karimipour}
\begin{center}
\maketitle{\it Institute for Studies in Theoretical Physics and Mathematics\\
P. O. Box 19395-5531, Tehran , Iran \\ Department of Physics , Sharif
Uinversity of Technology \\P O Box 11365-9161, Tehran , Iran \\
} \end{center}  \vspace {12 mm}

\begin{abstract}
We extend our previous analysis of the classical integrable models of Calogero in
several respects. Firstly we provide the algebraic resaons of their quantum integrability.Secondly we
show why these systems allow their initial value problem to be solved in closed
form .
Furthermore
we show that due to their similarity with the above models
the classical and quantum Heisenberg magnets with long rang
interactions in a magnetic field are also intergrable.
Explicit expressions are given for
the integrals of motion in involution in the classical case and for the commuting
operators in the quantum case.
\end{abstract}
\newpage
\section{Introduction}
Motivated by the discovery of Camassa, Holm, and Hyman [1] that the classical
system characterized by the Hamiltonian
\be H = \sum_{j,k = 1}^{N} p_j p_k exp [- \eta \vert q_j - q_k \vert ] \ee

with the usual poisson brackets is integrable, Calogero investigated the the
problem of integrabiltiy of the more general Hamiltonians
\be H = \sum_{j,k = 1}^{N} p_j p_k  f(q_j - q_k)\ee
with conserved quantities of the form \be c_{jk} = p_j p_k g ( q_j - q_k ) \ee
and succeeded to find a class of such systems [2], where \be f(x) = \lambda + \mu \cos ( x ) \ee
and \be g(x) = 1 - \cos ( x ). \ee ( Later he also investigated the classical [3 ] and quantum [4]
solvability integrability of other similar systems where the factor $p_j p_k $ appears
under a square root. ) It was shown in [5] that the integrability of systems of type (2,4)
is due to a very simple algebraic structure , i.e: the existence of N copies
of $ su(1,1) $ algebras one embeded inside the other, such that the Casimir operators
of all copies commute with each other. These Casimirs were shown to be the integrals
of motion of the system (2) which are in involution with each other. The purpose of this paper
which is a continuation of [5] is threefold. \\ 1) I extend
my previous considerations to the quantum case and show that quantum versions of these systems
and of more general systems are also
integrable .   \\
2 ) As a model with physical interest I study by algebraic methods the
integrability of classical and quantum Heisenberg magnets with long rang
interactions in a magnetic field .\\
3 ) It is known that integrability and " solvability in closed form" do not always
imply each other at least in practice. The models introduded in [2] have the
merit that they are integrable (i.e: possess N independent integrals of motion
in involuton ) and at the same time sovable (i.e: admit their initial value problem to be
solved in closed form ). Hence there must be a simple algebraic reason why
these models are also solvable in the above sence . We will provide the
 rational behind their solvability.

\section{Classical xxz Heisenberg Magnet}
The strategy that we follow is to take a system of classical vectors interacting
with each other and with an external field.
The dynamical variables
are given by a vector of length  squared equal to a constant $ C_1 $ .
\be {\bf S}_i = ( S_i^a ;  a = 1,2,3 \ \ \ {\bf S}_i\cdot {\bf S}_i = C_1 ) \ \ \ i =
1,2,... N  \ee
subject to the poisson bracket relations:
\be \{ S_i^a , S_j^b \} =  \epsilon ^{abc} S_j^c \delta _{ij} \ee
The Hamiltonian is given by :
\be H = \sum_{j,k = 1}^{N}  \lambda S^z_j S^z_k  +  \mu ( S^x_j S^x_k + S^y_j S^y_k ) + B \sum_{i=1}^N S_i^z  \ee
Mathematically the poisson bracket relations (7) are related with the Lie algebra
$ su(2)$ .  More presizely [6] the Lie algebra $ su(2 ) $ induces (7) as a natural poisson
bracket on it's dual which can be thought of as a Cartesian three dimensional space
with local coordiantes $ ( S^1 , S^2 , S^3 ) $. However this poisson bracket is degenerate
, since there are functions which poisson commute with everything , g.e:
$$ \{ {\bf S.S } , S^a \} = 0  $$
To obtain non-degenerate poisson bracket one should restrict oneself
to those submanifolds on which $ S \cdot S $ aquires a constant value $C_1$ i.e: the symplectic
leaves. \\
This system is general enough for our considerations. It will have several
subcases of interest:\\
{\bf case a) } $ S_i^x , S_i^y , $ and $ S_i^z $ are real with $ C_1 $  equal to a real number say one.
The symplectic leaves are two dimensional spheres. This is the classical Heisenberg Magnet
describing a system of classical spins interacting whit each other and with a
magnetic field B in the $ z $ direction.
For $ \lambda = \mu $ we have the isotropic magnet with $ su(2) $ symmetry.     \\
{\bf case b)} $ S_i^x $ and $ S_i^y  $  real and $ S_i^z $ pure imaginary with  $ C_1 = 0 $.
When $B = 0$, this is the system introduced by Calogero in [2] which is related to
the double cone symplectic leaf $$ {S^x}^2 + {S^y}^2 + {S^z}^2  = 0  $$ of the lie algebra $ su(1,1)$
. The canonical coordinates on this leaf are
\be S^z = ip \ \ \ \   S^x = p \cos q \ \ \   S^y = p \sin q \ee
and the Hamiltonian is given by (2,3) i.e:
\be H = \sum_{j,k = 1}^{N} \lambda p_j p_k + \mu \{ p_j p_k \ \cos [ ( q_j - q_k)] \}\ee
with the integrals of motion given by :
$$h_m =\sum_{j,k = 1}^{m}  p_j p_k ( 1 - \cos [ ( q_j - q_k)]   $$
and $$ P = \sum_{i=1}^{N} p_i $$

{\bf case b)} $ S_i^x $ and $ S_i^y  $  real and $ S_i^z $ pure imaginary with  $ C_1 \ne 0 $, .
This is the system introduced in [5] which is related to
the hyperboloidal symplectic leaf $$ {S^x}^2 + {S^y}^2 + {S^z}^2  = C_1  $$ of the lie algebra $ su(1,1)$
. The canonical coordinates on this leaf are
\be S^z = ip \ \ \ \   S^x ={ \sqrt {p^2 + C_1}}\cos q \ \ \   S^y = { \sqrt {p^2 + C_1}} \sin q \ee
and the Hamiltonian is given by :
\be H = \sum_{j,k = 1}^{N} \lambda p_j p_k + \mu \{ {\sqrt {p^2_j + C_1}}\ \
{\sqrt {p^2_k + C_1}}\ \ \cos [ ( q_j - q_k)] \} + B \sum_{j=1}^{j=N} p_j\ee
with the integrals of motion given by :
$$h_m =\sum_{j,k = 1}^{m}  p_j p_k - \{{\sqrt {p^2_j + C_1}}\ \ {\sqrt {p^2_k + C_1}}\ \ \cos [ ( q_j - q_k)] \}  $$
and $$ P = \sum_{i=1}^{N} p_i $$
All the results that we will obtain will apply to the above systems after minor
redefinitions of constants.

In order to understand in a systematic
manner the integrable structure of this system , we proceed as follows and define
the variables:
$$ X_1^a = S_1^a$$.......
\be X_m^a = S_1^a + S_2^a + . . . .  S_m^a \ee ......
$$ X_N^a = S_1^a + S_2^a + . . . . . . . . .  S_N^a $$
where
$ a \in ( 1\equiv x , 2\equiv y , 3\equiv z ) $

It is obvious that for each $m$ these sets of variables satisfy the same relations
among themselves as in (7) and form a copy of $ su(2) $ algebra , and furthermore since the smaller copies of the algebra
are embeded in the larger copies we have:
\be \{ X_m^a , X_n^b \} =  \epsilon ^{abc} X^c_{(m,n)}  \ee
where $ (m,n) $ is meant to denote the minimum of $m$ and $n$

Defining for each copy, say the $m$-th one the Casimir function \be  C_m = \sum_{a=1}^3 X_m^a X_m^a \ee
we obtain:
\be \{ C_i , X_j^b \} = 2 \epsilon ^{abc} X^a _i X^c_{(i,j)}  \ee
\be \{ C_i , C_j \} = 4 \epsilon ^{abc} X_i^a X_j^b X^c_{(i,j)}   \ee
We now note that in the last formula the indices $ i $ and $ j $ are not dummy variables
,however the index $ (i,j) $ is either equal to $i$
or to $j$ , in any case the tensor which is contracted with $\epsilon ^{abc}$
is symmetric with respect to the interchange of two of the indices $ (a,c) $ or $ (b,c) $ , hence the right hand side  identically vanishes:
$$\{C_i , C_j \} = 0 $$
It is interesting to note that although the Casimir of one copy  does not commute with the generators
of another copy as seen from (16) ,the Casimirs of different
copies commute among themselves.
However it should be noted from (16) that all the Casimir functions commute
with the generators of the largest copy. i.e:
\be \{ C_i , X_N^a \} = 0 \ee
The Hamiltonain can now be written as:
\be H =  \lambda Z_N^2 + \mu ( X_N^2 + Y_N^2 ) + B Z_N \ee or \be H = ( \lambda - \mu ) Z_N^2 + B Z_N +\mu C_N   \ee
It is seen that there are N integrals of motion in this system which are
in involution with each other and with the Hamiltonian. These are
\be I = \{ C_2 , C_3 , . . . .  C_N , Z_N \} \ee
We have found enough integrals to claim integrability of the
system. The explicit expressions of $ C_m $ are
\be C_m = \sum _{j,k = 1}^m S_j^x S_k^x + S_j^y S_k^y + S_j^z S_k^z \ee
and $ Z_N $ is the total spin ( Magnetization ) in the z-direction.\\
\section{ The Inintial Value Problem } In terms of the new variables one can easily write the equations of motions.
From (16) wee have :
\be \{ C_N , X_j^b \} = 2\epsilon ^{abc} X_N^a X_j^c \ee or
\be \{ C_N , X_j \} = 2 ( Z_N Y_j - Y_N Z_j ) \ee
\be \{ C_N , Y_j \} = 2 ( X_N Z_j - Z_N X_j ) \ee
\be \{ C_N , X_j \} = 2 ( Y_N X_j - X_N Y_j ) \ee

From which we can obtain the equations of motion .
\be  \left( \begin{array}{l} {d\over dt} X_j \\
{d\over dt} Y_j \\ {d\over dt} Z_j\\ \end{array} \right)
=  \left( \begin{array}{lll} 0 & { - B - 2\lambda Z_N }& { 2 \mu Y_N } \\
{  B + 2\lambda Z_N }& \ \ 0 & { - 2 \mu Y_N }\\
 {- 2 \mu Y_N } & { 2 \mu X_N }& 0
\\ \end{array} \right)
 \left( \begin{array}{l}  X_j \\
Y_j \\  Z_j\\ \end{array} \right)  \ee
which is of the form
\be {d\over {dt}}\Psi_j = \large A \Psi_j \ee
The time dependence of the matrix $A$ is itself determined from :
\be {d\over {dt}} X_N = \{ X_N , H \} = - ( 2(\lambda - \mu ) Z_N + B ) Y_N \ee
\be {d\over {dt}} Y_N = \{ Y_N , H \} =  ( 2(\lambda - \mu ) Z_N + B ) X_N \ee
\be {d\over {dt}} Z_N = \{ Z_N , H \} = 0  \ee
In any case $ Z_N $ is constant and the dynamics of $ X_N $ and $ Y_N $ depend on the
time dependence of the magnetic field. For a constant magnetic field the
components $ X_N $  and $ Y_N $ have a simple time evolution:
\be X_N = A \cos (\omega t + \alpha ) \ \ \ \  Y_N = A \sin (\omega t + \alpha ) \ \ \ \ee
where $ \omega = 2(\lambda - \mu)Z_N + B $ and $ A $ and $ \alpha$ are determined
from the initial conditions.
 Once the time dependence
of $ A $ is determined, the time dependence of $ \Psi_j $ will be determined
from (28):
\be  \Psi_j (t) = T exp \int_0^t A(t') dt'\Psi_j(0) \ee

However one can go beyond this formal solution. We note that $ A $ can be written
as follows:
\be A = - \omega L_3 - 2\mu ( X_N L_1 + Y_N L_2 ) \ee
where the matrices $ J_1 = iL_1 \ \  J_2 = iL_2 \ \ \ $ and $  J_3 = iL_3  $ are
the generators of rotation. Inserting (32) in (34) we find that :
\be A(t) = i \omega J_3 + i \mu A ( e^{ i( \omega t + \alpha ) } J_{-} + e^{ -i( \omega t + \alpha ) } J_{+} ) \ee
where $ J_{\pm} = J_1 \pm i J_2 $.
The dynamics of $ A $ is just a simple rotation around the z axis:
\be A = e^{ -i( \omega t + \alpha )J_3 } A(0) e^{ i( \omega t + \alpha )  J_{3}} \equiv  U^{-1} (t) A(0) U(t) \ee
Multiplying both sides of this equation by $ U(t) $  and defining $ \Phi _i (t) =  U (t) \Psi_i (t) $ we find:
$$ {d\over dt} \Phi _i(t) = ( A(0) - {d\over dt}U(t) U(t) ) \Phi _i (t) $$
\be = ( A(0) - i  \omega  J_3 ) \Phi _i (t) \ee
This is a simple evolution equation governed by a constant matrix the solution of which is simply
$$ \Phi _i(t) =  e ^ {( A(0) - i \omega J_3 )t} \Phi _i (t) $$
We have thus arrived at a closed form  solution of the initial value problem for $ \Phi _i $ and thus for $ \Psi_i $.
Our analysis shows clearly why the initial value problem in [2] can be solved in closed form.
\section{ The Quantum Case }

In the quantum case the field variables $ S_i^a $ are replaced by spin operators
acting on a Hilbert space $ \large V = V^{\otimes N } $ where $ V $ is an irreducible representation
space of $ su(2) $.( we do not restrict ourselves to the spin $ {1\over 2 } $ representation
and :
$$ s_i^a = 1\otimes 1\otimes . . . .  1\otimes s^a \otimes . . . .  \otimes 1 $$
Where $ s^a $ acts on the i-th space . The poisson brackets are replaced by
\be [ S_i^a , S_j^b ] = i \epsilon ^{abc} S_j^c \delta _{ij} \ee
and the hamiltonian has the same form as before. The role of symplectic
leaves $ {\bf S.S } = C_1 $ is now played by a particular irreducible representation
where the Casimir operator takes a constant value.
Again for discussion of integrability we define operators
\be X_m^a = s_1^a + s_2^a + . . . .  s_m^a \ee
Exactly as in the classical case we obtain
\be [ X_m^a , X_n^b ] = i \epsilon ^{abc} X^c_{(m,n)}  \ee
Note that each operator $ X_m^a $ say $ Z_m $ acts like the third component of total
spin operator in the first $ m $ spaces and acts trivially in the rest of
spaces.
Defining now the operators
$$ C_m = X_m^a X_m^a $$ ( where a sum over the index $ a $ is understood )
we obtain
\be [ C_m , X_n^b ] = i \epsilon ^{abc} ( X^a_m {X^c}_{(m,n)} + {X^c} _{(m,n)} X^a_m )  \ee
\be [ C_i , C_j ] = i \epsilon ^{abc} \bigg (  X_n^b \big ( X_m^a {X^c}_{(n,m)} + {X^c} _{(n,m)} X_m^a \big ) +
\big ( X_m^a {X^c}_{(n,m)} + {X^c} _{(n,m)} X_m^a \big ) X_n^b \bigg )  \ee
Again we note that in the last formula the indices $ i $ and $ j $ are not dummy variables
,however the index $ (m,n) $ is either equal to $m$
or to $n$ , in any case the tensor which is contracted with $\epsilon ^{abc}$
is symmetric with respect to the interchange of two of the indices $ (a,c) $ or $ (b,c) $ , hence the right hand side  identically vanishes:
$$[ C_m , C_n ] = 0 $$
It should be noted from (41) that all the Casimir operators commute
with the generators of the largest copy. i.e:
\be [ C_m , X_N^a ] = 0 \ee
The family of commuting operators is the following set:
\be I = \{ C_2 , C_3 , . . . .  C_N , Z_N \} \ee
Where the explicit expression for $ C_m $ the same as in (22).
{\bf Remark } :None of the formulas and results in this section depends on which
representation of $ su(2) $ sits on different sites . They are also independent
of the particular algebra which we use . In fact what we have found is true
for any irreducibe representation of any simple Lie algebra, provided
that one adds to the set I all the higher order Casimir operators of the
algebra.

\section{Discussion}
We have provided a mathematical basis in which the integrability
and solvability of the models of [2] and their generalizations
is explained.
A very interesting problem is to find a mathematical formalism in which the
integrability of systems like (1) whose Hamiltonians are not factorized
can be explained. However the systems studied in [2]
are perhaps a good testing ground and starting point for studying
integrability in systems with local interactions. This may reminds us of the
Mean Field Method which we use when we first encounter a new statistical system.
It may be appropriate to call such systems Mean Field Integrable Models.

\section{Acknowledgments} I would like to thank M.R. Daj , M.R Ejtehadi, R.Golestanian,
V. Milani, M.R. Rahimi Tabar and S. Rouhani for very stimulating discussions.

\newpage
{\large \bf References}
\begin{enumerate}

\item  R. Camassa, and D.D. Holm , Phys. Rev. Lett. 7(1993)1661;
R.Camassa,D.D. Holm, and J.M.Hyman, Adv.Appl.Mech. 31(1994)1
\item  F. Calogero : Phys. Lett. A. {\bf 201} (1995) 306-310
\item  F. Calogero : Jour. Math. Phys. {\bf 36} 9 (1995)
\item  F. Calogero and J.F. van Diejen : Phys. Lett. A. {\bf 205} (1995) 143
\item  V. Karimipour " Algebraic and Geometric Structure of the Integrable Models
recently proposed by Calogero, IPM preprint Feb.96 , Tehran.
\item  L. D. Faddeev ; Integrable Models in 1+1 dimensional quantum Fields
theory
(Les Houches Lectures 1982), Elsevier, Amsterdam (1984)
\end{enumerate}
\end{document}